\documentclass[aps,showpacs,amsmath,amssymb,longbibliography, notitlepage,superscriptaddress, twocolumn]{revtex4-2}
\usepackage{floatrow}
\usepackage{graphicx}
\usepackage{amssymb}
\usepackage{amsmath}
\usepackage{bm}
\graphicspath{{./imgs/}}
\usepackage{textcomp}
\usepackage{wrapfig}
\newcommand{\ts}{\textsubscript}
\usepackage{tikz}

\newcommand{\appropto}{\mathrel{\vcenter{
  \offinterlineskip\halign{\hfil$##$\cr
    \propto\cr\noalign{\kern2pt}\sim\cr\noalign{\kern-2pt}}}}}
\DeclareRobustCommand{\circledno}[1]{%
  \tikz[baseline=(char.base)]{
    \node[shape=circle,draw,inner sep=1pt] (char) {\scriptsize #1};
  }%
}

\begin{document}

\title{Stochastic elastohydrodynamics of soft valves}

\author{Mengfei He}
\affiliation{School of Engineering and Applied Sciences, Harvard University, Cambridge MA 02138.}
\author{Sungkyu Cho}
\affiliation{Boston Children's Hospital, Boston, MA 02115, USA}
\author{ Gianna Dafflisio }
\affiliation{Boston Children's Hospital, Boston, MA 02115, USA}
\author{Sitaram Emani}
\affiliation{Boston Children's Hospital, Boston, MA 02115, USA}
\author{L. Mahadevan}
\email{lmahadev@g.harvard.edu,\\sitaram.emani@cardio.chboston.org}
\affiliation{School of Engineering and Applied Sciences, Harvard University, Cambridge MA 02138.}
\affiliation{Department of Physics, Harvard University, Cambridge MA 02138.}
\affiliation{Department of Organismic and Evolutionary Biology, Harvard University, Cambridge 02138}

\begin{abstract}
Soft valves serve to modulate and rectify flows in complex vasculatures across the tree of life, e.g. in the heart of every human reading this. Here we consider a minimal physical model of the heart mitral valve modeled as a flexible conical shell capable of flow rectification via collapse and coaptation in an impinging (reverse) flow.  Our experiments show that the complex elastohydrodynamics of closure features a noise-activated rectification mechanism. A minimal theoretical model allows us to rationalize our observations while illuminating a dynamical bifurcation driven by stochastic hydrodynamic forces. Our theory also suggests a way to trigger the coaptation of soft valves on demand, which we corroborate using experiments, suggesting a design principle for their efficient operation.
\end{abstract}

\maketitle

\begin{figure}[htb]
\centering
\includegraphics[width=0.95\textwidth]{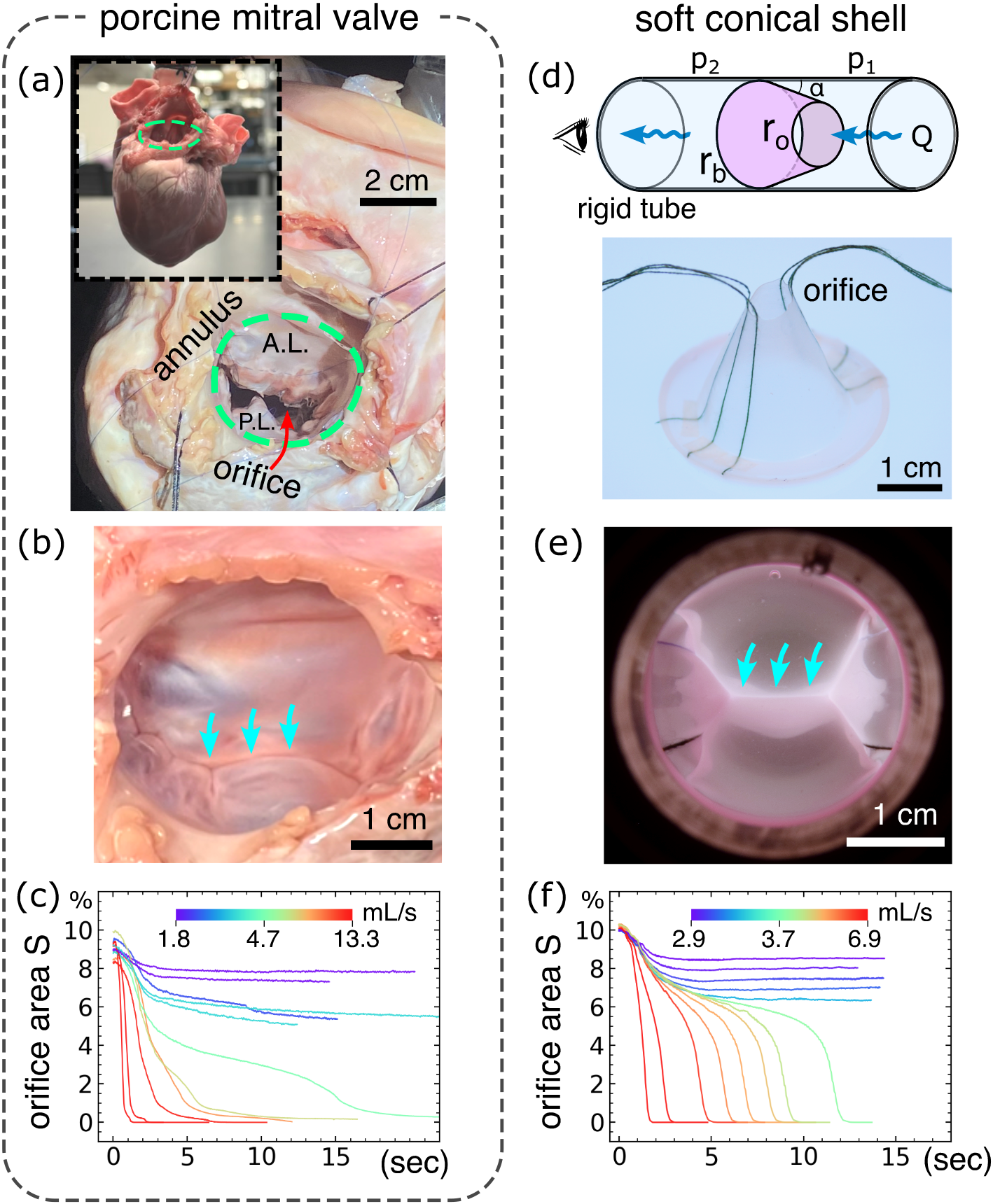}
\caption{A tethered soft cone mimic of a heart mitral valve. 
(a) Porcine mitral valve viewed from a Yorkshire pig heart with its left atrium excised.  A. L./P. L., anterior/posterior leaflets.  
(d) Tethered soft conical shell made from Vinylpolysiloxane silicone elastomer, placed under flow.
(b) and (e) Closed states of both valves.  Arrows: coaptation lines.
(c) and (f) Orifice size as a function of time, at various flow rates for both valves.
}
\label{fig:porcine_inspired}
\end{figure}

Flow rectification in many engineered systems, from microelectronics and fluidics  to artificial neurons work using actively controlled valves and are the basis for many modern technologies that require the directional propagation of information, matter and energy. 
However, rectifiers in biological systems, such as valves in the heart and lymphatic and venous systems, work without active control. Instead, their function emerges spontaneously from the interplay between flexibility and flow~\cite{peskin82,oliver2004,heil2011}. These valves are inherently three-dimensional, function in high-Reynolds number flows, must withstand variations due to growth, individual differences, and environmental fluctuations, and also last for a lifetime.  
As an example, in Fig.~\ref{fig:porcine_inspired}a, we show a porcine mitral valve consists of a two-lobed sheet that forms a dynamic orifice to facilitate forward blood flow. 
When a critical reversed flow rate or adverse pressure gradient is reached, the valve lobes seal tightly (Fig~\ref{fig:porcine_inspired}b), forming a coaptation state that prevents backflow. Complementing the many studies on the anatomical and dynamical complexity of the heart valve using large-scale computation studies~\cite{peskin82, sotiropoulos16, toma16, kaiser19, lee2021}, here we consider a physical mimic to understand the dynamics of these soft valves using simple experiments and a minimal theory.

Inspired by the mammalian mitral valve, we mold thin conical shells of base and orifice radii $r\ts{b}, r\ts{o}$, and opening angle $2\alpha$, from Vinylpolysiloxane elastomer (VPS, Zhermack 8 Shore A), with and without embedded chordae.
The shell is then mounted in an acrylic tube with the same radius as the cone base to form a testing module (Fig.~\ref{fig:porcine_inspired}d).
The flow rate, $Q$, and the pressure difference, $p\equiv p_1-p_2$, across the shell are monitored with a turbine flow meter (Vision Turbine Meters BV2000) and a differential pressure transducer (Validyne P55D), respectively. De-ionized water is driven from a far-field pressure difference from two end reservoirs (4 L) to flow through the testing module, regulated by a flow controller (Elveflow OB1 MK4). 
High-speed imaging (Phantom v9.1) allows us to record the dynamics of the shell (with and without chordae) subject to flow in either direction.

\begin{figure*}[bt]
\includegraphics[width=0.95\textwidth]{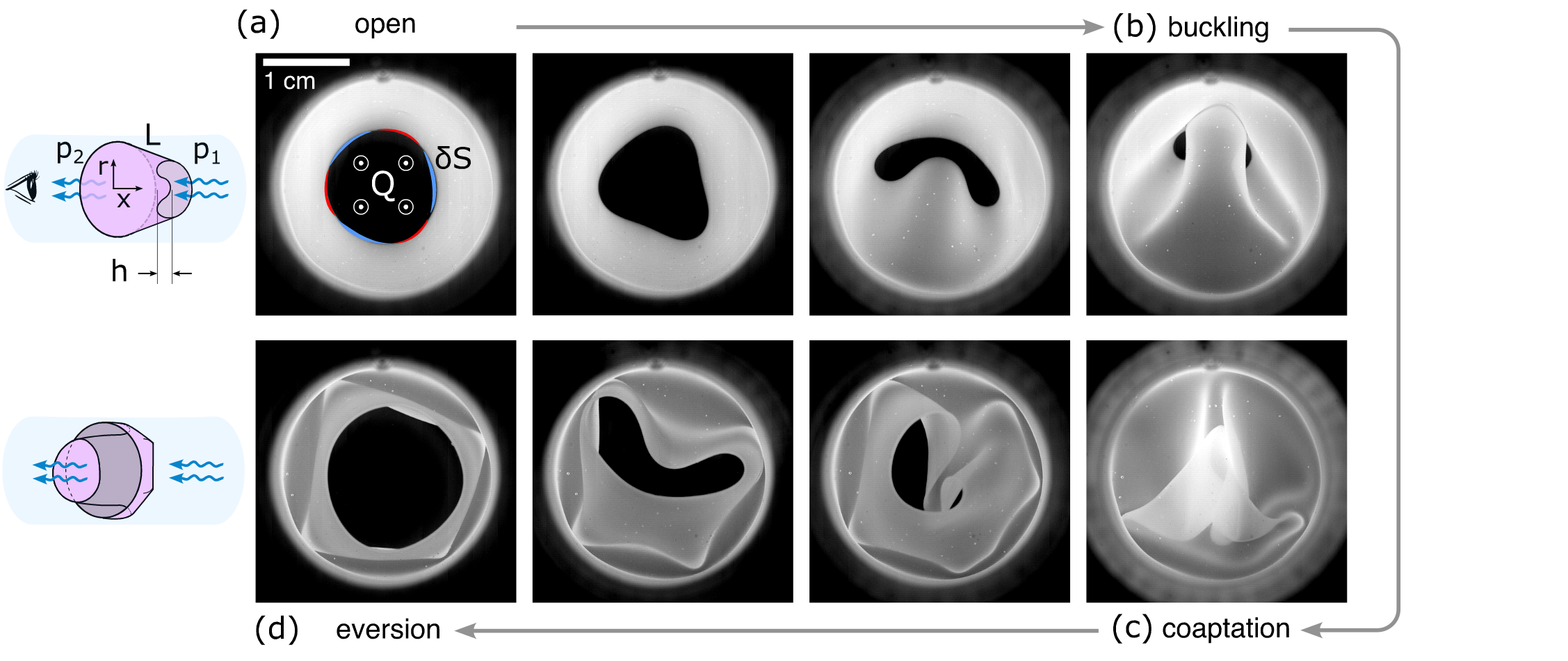}
\caption{ 
High-speed images of a conical shell ($\alpha=10^{\circ}$) buckling and eversing under an impinging flow ($Q=11.5$ mL/s), viewed from the back of the cone.  The shell first oscillates (a) before buckling (b) to form a coaptation (c).  The shell eventually everses (d).  The schematics show the orientations of the initial and the eversed cone.
 }
\label{fig:buckle}
\end{figure*}

When subject to an adverse pressure gradient in the direction opposing the narrowing, a tethered cone ($\alpha=20^{\circ}$, Fig.~\ref{fig:porcine_inspired}d) closes to form a two-leaflet coaptation with a similarly branched coaptation line to that seen in the mitral valve (Fig.~\ref{fig:porcine_inspired}b, e). Measuring the time evolution of orifice size under adverse pressure reveals behaviors akin to that observed in the mitral valve (Fig.~\ref{fig:porcine_inspired}c, f): the orifice deforms gently (purple to cyan) when the adverse pressure gradients are small, but when the gradient exceeds a threshold, the valve closes rather suddenly (cyan to red). 


To closely examine the onset of valve closure, we use a high-speed camera to record the morphology of an untethered soft conical shell ($\alpha=10^{\circ}$) under an impinging flow ($Q=11.5$ mL/s), 
shown in Fig.~\ref{fig:buckle}a-d.
At a low flow rate, a small oscillation is excited at the cone's orifice, sweeping out area $\delta S$ during each cycle (red/blue regions, Fig.~\ref{fig:buckle}a)\cite{lee2021}.
As the flow rate increases, the oscillation amplitude of one lobe grows at the cost of the others, causing it to buckle inwards (Fig.~\ref{fig:buckle}b).  
The cone then collapses into a state of \emph{coaptation}, wherein the shell self-contacts to form a seal that prevents further reverse flow (Fig.~\ref{fig:buckle}c).  
Increasing the flow even further eventually forces the shell to evert, causing valve failure through opening and subsequent flow reversal (Fig.~\ref{fig:buckle}d).

\begin{figure*}[bt]
\centering
\includegraphics[width=1\textwidth]{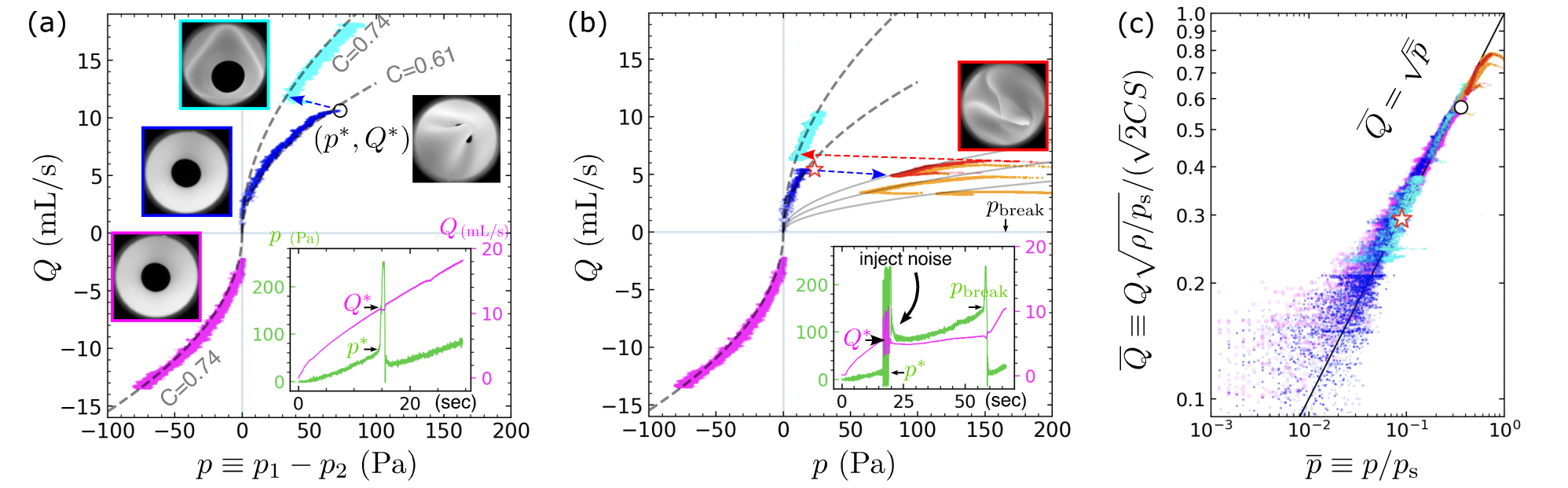}
\caption{  
Pressure-flow ($p$-$Q$) relation for impinging flow on a conical shell valve.
(a) Under a near steady flow, the flow rate $Q$ through the shell as a function of the pressure ($p\equiv p_1-p_2$), for the flow along (magenta) or against (blue) the direction of narrowing, and for an eversed cone (cyan) post buckling at $(p^*, Q^*)$ (black open circle). Upon buckling, the system transitions abruptly from the buckling point (black open circle) to the eversion branch (cyan), leaving a gap with no steady state.
(b) Injected flow noise causes an early buckling transition (red open star), resulting in a coaptation branch with a small leak (red).  Sequential folding states can further split the coaptation state into sub-branches (orange). $p\ts{break}$: threshold pressure for the eventual cone eversion (cyan).
Inset images: high-speed photographs of the corresponding orifice shape for each branch.
Inset plots: time series of the pressure $p$-$t$ (green) and the flow rate $Q$-$t$ (magenta).
Dashed and solid curves: theoretical prediction Eq.~\ref{eq:pQ} for an inviscid flow, with measured values of the effective coefficient of contraction, $C$.
(c) Rescaled pressure-flow relation, $\overline{p}$-$\overline{Q}$, using the Seide buckling load, $p\ts{s}$, collapsing all branches in (a), (b).  Solid line: $\overline{Q}=\sqrt{\overline{p}}$.  Under an impinging flow the coaptation transition (black open circle) occurs near $(\overline{p}, \overline{Q}) = (1, 1)$, while early buckling under noise (red open star) occurs an order of magnitude lower.
}
\label{fig:pQ}
\end{figure*}

Gradually ramping up the far-field driving pressure, we simultaneously track the \emph{local} pressure drop across the cone $p\equiv p_1-p_2$, and the flow rate, $Q$, as a function of time, $t$. In Fig.~\ref{fig:pQ}a (inset) we show that both $p$ (green markers) and $Q$ (magenta markers) steadily increase until, at the thresholds of $p^*$ and $Q^*$, the cone buckles and  everts, leading to a decrease in $p$ preceded by a sharp spike. Synchronizing the time series for pressure and flow rate allows us to examine $Q$ as a response to $p$ as depicted in Fig.~\ref{fig:pQ}a. The system first follows a smooth trajectory (blue markers). When the system reaches the threshold, $(p^*, Q^*)$ (black open circle), there is a discontinuous transition (blue dashed arrow) toward the eversion branch (cyan markers).
On the other hand, when we reverse the far-field driving ($Q<0$, forward flow along the direction of cone narrowing), the cone deforms minimally, and the $p$-$Q$ curve exhibits a simple trend (magenta markers, III quadrant).  We show corresponding high-speed images of the cone next to each branch.  To rationalize the pressure threshold for the buckling, we note that for a conical shell~\cite{seide1956}, a rough estimate for the force is given by  $f\ts{s}=4\pi\sqrt{YB}\cos^2\alpha$, where $Y$ and $B$ are the stretching and bending moduli, so that the pressure is $p\ts{s} = f\ts{s}/(r\ts{b}^2-r\ts{o}^2)\approx204$ Pa, in agreement with the order of magnitude of $p^* \sim 10^2$ Pa obtained in our experiment.

It is not surprising that the buckling transition occurs below the classical buckling limit, given the inevitable imperfections of the system and the well-known subcritical nature of shell buckling~\cite{koiter, thompson2015}.  Flow visualization using polyamide seeding particles further confirms significant fluctuations in flow (see SI).  To probe the effect of these imperfections, we actively added noise. The inset of Fig.~\ref{fig:pQ}b shows that a short duration of noise at $\sim 18$ s, injected by a syringe connected in front of the cone, causes the cone to collapse at much smaller $p^*$, $Q^*$. 
In such cases, we see an extended state of coaptation with a small leak through small incompletely coapted pleats ($\sim18$ s to $\sim 60$ s~\footnote{We further decrease the rate of ramp of the far-field driving pressure to highlight the extended state of coaptation in this case}, inset of Fig.~\ref{fig:pQ}b).
Consequently, the $p$-$Q$ trajectory shown in Fig.~\ref{fig:pQ}b reaches a threshold earlier (red open star), transitioning discontinuously (blue dashed arrow) to the coaptation branch, where the flow rate remains low (red markers). Further increase in the pressure leads to secondary foldings that cause secondary coaptation branches as the crumpled cone jumps among new states (orange markers). Overall, early transition induced by noise facilitates earlier onset of flow \emph{rectification}, leading to a differential resistance between impinging (blue/red) and forward (magenta) flows over a much wider pressure
range than in the absence of noise.
Eventually, when the pressure exceeds a second, much higher threshold $p\ts{break}$, the cone everts fully (red dashed arrow)  to a new state (cyan markers), analogous to the reverse breakdown that is well known in semiconductor diodes.

A unified picture of the parabolic $p$-$Q$ curves follows from Bernoulli's equation which yields
\begin{align}
Q = \sqrt{2}CS\sqrt{\frac{p}{\rho}},
\label{eq:pQ}
\end{align}
where $C<1$ is the coefficient of contraction~\cite{birkhoff}, and $S$ is the cross-sectional area of the orifice.  Interestingly, even though Eq.~\ref{eq:pQ} is strictly valid only for inviscid  flows,  treating $C$ as an effective contraction coefficient allows us to capture the effects of \emph{fluctuating} flows past both rigid and flexible cones at large Reynolds numbers ($\sim 10^3$) (see SI for details),
for the different $p$-$Q$ branches in Fig.~\ref{fig:pQ}a, b (dashed and solid curves). Rescaling $p$, $Q$ with the estimate of the buckling pressure~\cite{seide1956} leads to $\overline{p} = p/p\ts{s}$ and $\overline{Q} = Q\sqrt{\rho/p\ts{s}}/(\sqrt{2}CS)$, collapsing all $\overline{p}$-$\overline{Q}$ curves into a single one, as shown in Fig.~\ref{fig:pQ}c. Furthermore, we note that the coaptation transition occurs near the classical limit $(\overline{p}, \overline{Q})=(1,1)$ for a steady flow (black open circle), but almost an order of magnitude lower for a noisily driven flow (red open star), suggesting a natural way to control the transition using noise.

\begin{figure*}[t]
\centering
\includegraphics[width=0.98\textwidth]{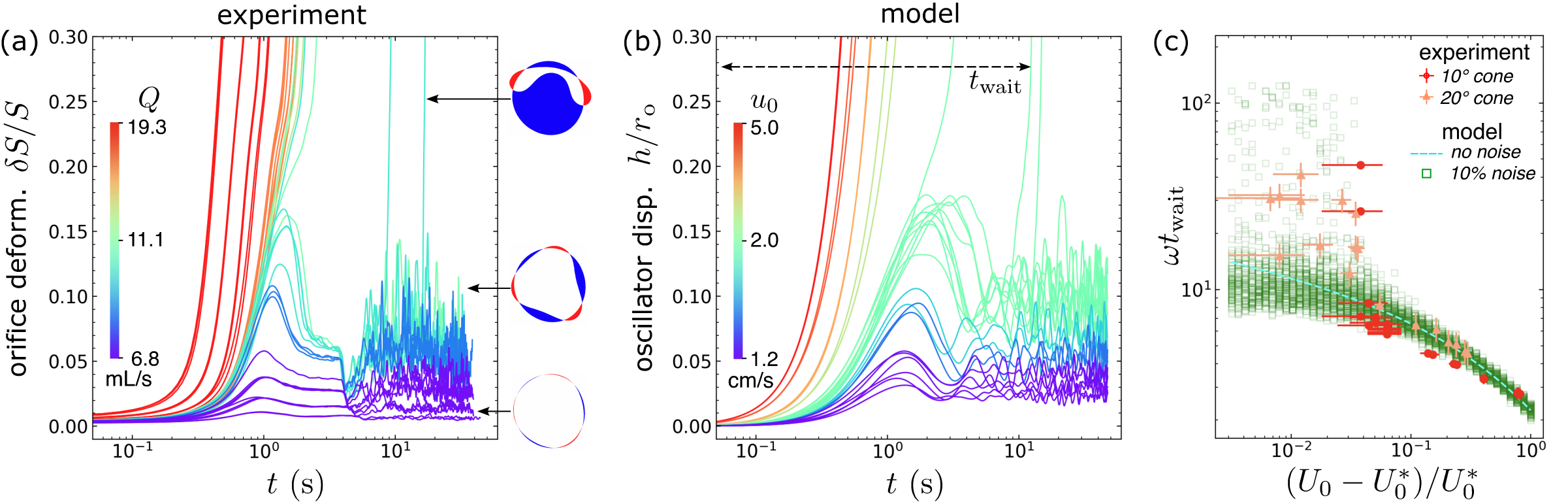}
\caption{Comparing experiment and theory for the flow-induced coaptation transition.
(a) Measured orifice deformation, $\delta S/S$, of a cone ($\alpha=10^{\circ}$) under various flow rates (colorbar). 
(b) Time evolution of the displacement $x\equiv h/r\ts{o}$ from the stochastic nonlinear oscillator model (Eq.~\ref{eq:oscillator}) (see SI for parameter values) and various flow velocities (colorbar). $t\ts{wait}$: the waiting time for the oscillator to diverge.
(c) The normalized waiting times, $\omega t\ts{wait}$, for two different cones to buckle (red and salmon markers, $\omega=2.66$ and $6.45$ s$^{-1}$ respectively) and that for the model oscillator to diverge (green markers), all plotted as a function of the normalized distance to the threshold, $(U_0-U_0^*)/U_0^*$.  Cyan dashed curve: the waiting times given by the model with no flow noise.}
\label{fig:dynamics}
\end{figure*}


To probe the stochastic nature of the cone buckling transition in flow, we measure the orifice area variation, $\delta S$, as a function of time at various flow rates (color bar) using high-speed imaging~\footnote{We have verified that other definitions of the orifice deformation give essentially the same result, while the area measure adopted here, which is defined as the total area of the blue and the red regions, has the advantage of robustness and ease of computation.}.
In Fig.~\ref{fig:dynamics}a we show that at low flow rates the orifice first contracts, producing an initial bump in the $\delta S$-$t$ curve before entering into an oscillating state (purple to blue). Both the mean and the amplitude of the oscillation increases as $Q$ increases.  Near the threshold $Q^*$, the fluctuation amplitude becomes large enough to collapse the cone, indicated by the diverging curves of $\delta S$ (cyan).  At an even higher flow rate $Q > Q^*$, the orifice closes upon the impinging flow deterministically without entering the fluctuating state (green to red).

A minimal picture of the initiation of cone buckling is afforded in terms of the dominant mode of the small oscillation via the relation $\delta S \sim  2\pi r\ts{o}h$, where $h(t)$ is the displacement of the orifice edge around its rest shape, $r=r\ts{o}$.  Under this 1-dimensional representation of the deformation of a cone with length $L$, thickness $e$ and Young's modulus $E$ (see schematic of Fig.~\ref{fig:buckle}), we may write an elastohydrodynamic equation of motion for $h(t)$ as ~\cite{argentina05,mandre2010}:
\begin{align}
\rho L h_{tt}+f\ts{e}(h)+Dh_t-\Delta P=0.\,
\label{eq:elastohydrodynamics}
\end{align}
The first term reflects the added-mass inertia associated with displacing the surrounding fluid of density $\rho$ by the cone \footnote{Since the shell thickness $e\ll L$, we neglect the mass of the cone  relative to the mass of fluid displaced since $\rho_s e \ll \rho L$}. The second term $f\ts{e}=\rho L \omega^2(h - h^2/\ell)$ follows from a Landau-like expansion of the elastic restoring stress incorporating geometric nonlinearities (we note that a conical shape implies that $h \rightarrow -h$ is not a symmetry). Here, the characteristic frequency $\omega$ is determined by balancing the fluid inertia and the elastic bending~\cite{virot2020}
$\omega\sim \sqrt{E/\rho}e^{3/2}r_\ts{o}^{-2}L^{-1/2}$.

The third term corresponds to a linear damping where $D\equiv \rho L \cdot 2 \zeta \omega$ with $\zeta$ the dimensionless damping ratio. 
The last term in Eq.~\ref{eq:elastohydrodynamics} is the hydrodynamic pressure of the impinging fluid, whose leading term scales as~\cite{argentina05} 
$\Delta P=\rho u[\partial_t h+ u(\partial_x h+\alpha)]$.
Here, $u=u_0+\xi$ is the flow velocity, which we decompose into a base flow, $u_0$, and a persistent noise of normal distribution of zero mean $\xi(t)$. 
Scaling the variables using the definitions $x\equiv h/r\ts{o}$, $\lambda\equiv\ell/r\ts{o}$, $\tau\equiv t\omega$, $U = U_0+\Xi \equiv(u_0 + \xi)/(\omega L)$, and $\overline{\alpha}\equiv\alpha L/r\ts{o}$ in Eq.~\ref{eq:elastohydrodynamics}, we obtain a minimal model for a stochastic driven-damped nonlinear oscillator:

\begin{align}
\label{eq:oscillator}
\frac{\mathrm{d}^2x}{\mathrm{d}\tau^2}+(2\zeta-U)\frac{\mathrm{d}x}{\mathrm{d}\tau}-\frac {1}{\lambda}x^2 + (1-U^2)x - U^2\overline{\alpha}=0.
\end{align}
We note that this corresponds to the normal form for an inertial noisy saddle-node bifurcation~\cite{kuehn2015} producing two equilibria corresponding to a saddle and a node at $x_{\pm}=(\lambda/2)[1-U_0^2\pm\sqrt{(1-U_0^2)^2-4U_0^2\overline{\alpha}/\lambda}]$ that coalesce as the (dimensionless) driving velocity, $U_0$, is increased, leading to a bifurcation.  The threshold velocity, $U_0^*(\zeta,\lambda,\overline{\alpha})\equiv u_0^*/(\omega L)$, is given by enforcing the system's initial condition onto the boundary of the attraction basin (separatrix) of the stable focus $x_-$ (see SI for details).

With parameters measured from the cone shown in Fig.~\ref{fig:dynamics}a (see SI), numerical integration of Eq.~\ref{eq:oscillator} (with $\sqrt{\langle\Xi^2\rangle}/U_0 = 10^{-1}$) yields the results shown in Fig.~\ref{fig:dynamics}b. Small values of $U_0$ push the oscillator toward the saddle point, $x_{+}$, but not beyond it, leading to an oscillating recovery to the focus $x_{-}$, similar to that seen experimentally in Fig.~\ref{fig:dynamics}a (purple).
When $U_0$ is increased beyond the threshold $U^*_0$, the oscillator overshoots the saddle point and diverges (red).  In the marginal case $U_0\rightarrow U_0^*$,  the oscillator fluctuates with a large amplitude before it diverges stochastically (cyan). 

To characterize the role of the fluctuations on the onset of rectification, we use our minimal oscillator model to record the waiting time $t\ts{wait}$ for the response to diverge following flow initiation at $t=0$, as a function of the mean flow velocity $U_0$.  
Figure~\ref{fig:dynamics}c shows that compared to the deterministic limit (cyan dashed curve),  persistent noise $\sqrt{\langle\Xi^2\rangle}/U_0 = 10^{-1}$ gives rise to a spread of $t\ts{wait}$, which further widens as $U_0 \rightarrow U_0^*$.
In the same space, we plot the experimentally measured waiting times for two separate conical shells with opening angles $\alpha=10^{\circ}$ (red) and $\alpha=20^{\circ}$ (salmon).  When normalized by the characteristic frequency $\omega$, these measurements collapse well, agreeing quantitatively with the simulation results of the model. Not surprisingly, our model leads to an extreme value (Frechet) distribution for waiting times (see SI for details).

\begin{figure}[htp]
\centering
\includegraphics[width=0.85\textwidth]{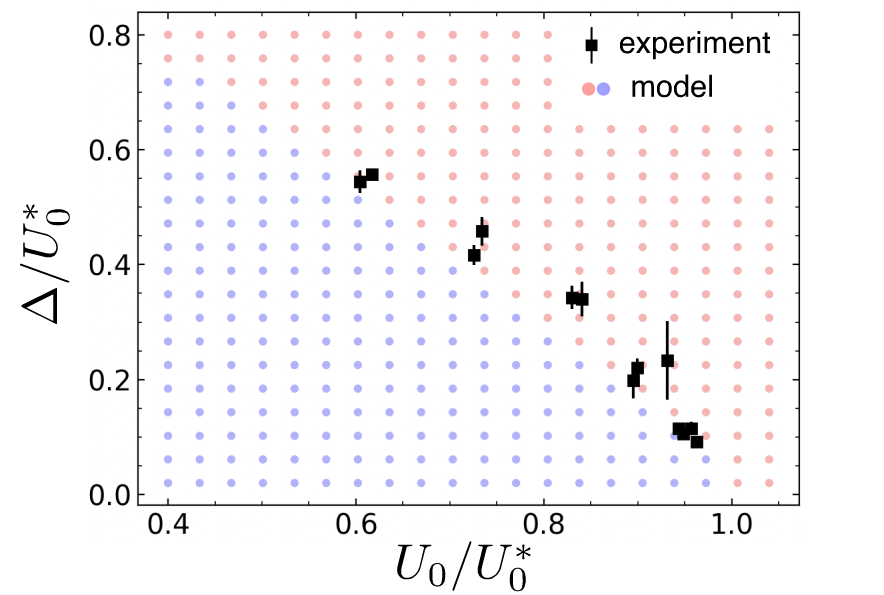}
\caption{Controlling coaptation and closure. The noise amplitude $\Delta$ required to collapse a cone ($\alpha=20^{\circ}$) as a function of the base flow $U_0$, both normalized by the threshold $U_0^*$. Blue (red) markers: converging (diverging) terminal states given by Eq.~\ref{eq:oscillator}, after a disturbance of amplitude $\Delta$. Black markers: measured minimal disturbance spike required to collapse a cone under an impinging flow.}
\label{fig:spikes}
\end{figure}

Having experimentally seen that our soft valve can be modeled as a stochastic nonlinear oscillator near a \emph{subcritical} bifurcation, we ask: Can we trigger a recification bifurcation on demand (rather than just via increasing the noise variance), even when the pressures are below the critical threshold? To probe this, we consider the effect of a disturbance of a prescribed amplitude,  replacing the persistent random drive $\xi$ with a single spike
$\Delta e^{-\tau^2/2\Sigma^2}$ (we set the dimensionless spike width $\Sigma=\omega \times 0.5$ s to match the typical experimental noise spectrum), and scan the base-noise $U_0$-$\Delta$ space using Eq.~\ref{eq:oscillator}. In Fig.~\ref{fig:spikes}, we see a well-defined boundary between the diverging (red) and converging (blue) end states of the oscillator when $\tau\rightarrow\infty$. To compare with the experiment, we use the syringe connected with our setup to inject a sequence of spikes of progressively larger amplitude, sufficiently separated to avoid cumulative effects, until the cone collapses (see SI for details). In the $U_0$-$\Delta$ space, the measured final spikes that collapse the cone for each flow rate scatter tightly around the simulated phase boundary.

The dynamical behavior of soft valves as flow rectifiers is potentially complicated by the interplay between geometry, nonlinearity and stochasticity, which suggests that the only recourse is elaborate experiments and/or large-scale simulations. Here we have shown that we can distill the essential mechanisms at work using a simple experiment.  Our study of a conical shell model subjected to an impinging flow has uncovered two distinct post-buckling states: in near-laminar conditions, the cone flips through at a threshold flux into a dysfunctional, everted state, whereas in the presence of disturbances, the cone buckles dynamically and transitions into an extended functional coaptation state, and \emph{noise} enhances rectification in the soft valve system. A minimal theory allows us to quantitatively predict the stochastic transition and the waiting time for this buckling onset. We also show that we can use controlled disturbances to trigger the rectification transition on demand, thus naturally providing a means to guide it dynamically.  This suggests a role for active mechanisms associated with myocardium or papillary muscle contractions in the real heart valve that can act to tune the passive processes at play. Natural questions for future study include adding a periodic forcing on the flow to recapitulate biological reality, as well as accounting for the forces from the active chordae and the soft walls of the heart. 

\textit{Acknowledgements.}We thank the Simons Foundation, the Henri Seydoux Fund, the TreeFrog Fund and Boston Children's Hospital for partial financial support.

%

\vspace{3cm}

\renewcommand{\thefigure}{S\arabic{figure}}
\setcounter{figure}{0} 
\renewcommand{\theequation}{S\arabic{equation}}
\setcounter{equation}{0}
\renewcommand{\thesection}{SI \arabic{section}}
\setcounter{section}{0}

\noindent\textbf{\large Supplementary Material for ``Stochastic elastohydrodynamics of soft valves"} \\
\noindent\textbf{\footnotesize Mengfei He, Sungkyu Cho, Gianna Dafflisio, Sitaram Emani and L. Mahadevan}

\section{Experimental methods}

\begin{figure}[htbp] 
\includegraphics[width=0.9\textwidth]{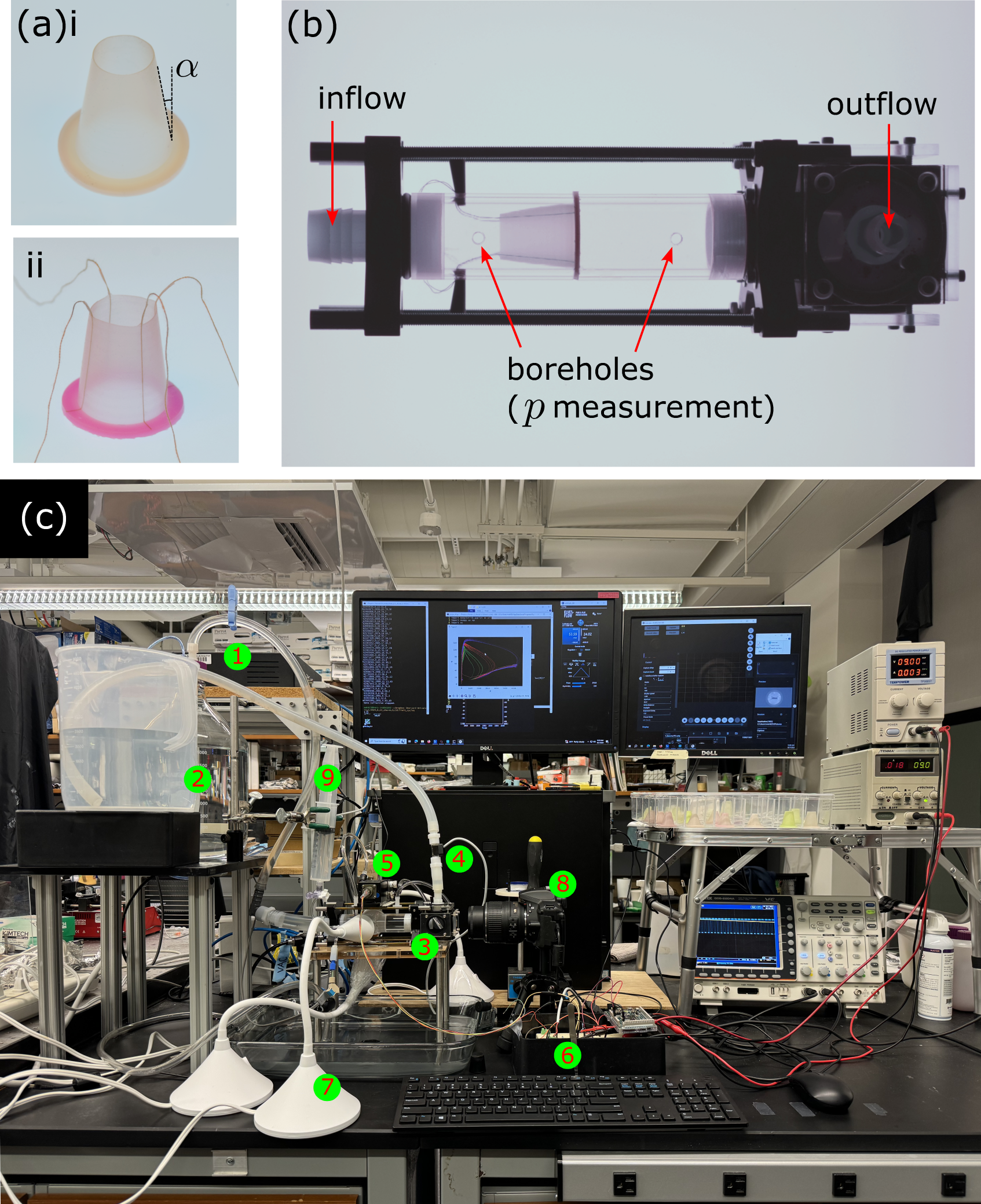}
\caption{(a)i Bare VPS conical shell of half opening angle $\alpha=10^{\circ}$. ii VPS shell of $\alpha=7^{\circ}$ with four embedded cords. (b) The testing module. (c) Setup overview.
\circledno{1} flow controller,
\circledno{2} fluid reservoir, 
\circledno{3} testing module, 
\circledno{4} flow meter, 
\circledno{5} pressure transducer, 
\circledno{6} operational amplifier, 
\circledno{7} illumination, 
\circledno{8} camera, and 
\circledno{9} syringe.}
\label{fig:setup}
\end{figure}

We fabricate thin conical shells from a Vinylpolysiloxane silicone elastomer (VPS, Zhermack 8 shore A).  Base and curing agents of VPS are mixed at a weight ratio of 1:1 before being placed in a vacuum desiccator (4 minutes, 25$^{\circ}$C) to remove microbubbles.  The precursor fluid is then gapped between two conical molds and cured (20 minutes) into a thin conical shell.  The opening angle $\alpha$ of the elastic cone is controlled by the mold geometry, while the shell thickness is set by the gap size between the molds.  Sewing threads are cured into the conical shell to mimic the functionality of the system of \textit{chordae tendineae} (heart strings), which confines the motion of the shell membrane to prevent prolapse.  We truncate the conical shell with a razor blade before peeling it off the mold. Completed samples, with and without embedded threads, are shown in Fig.~\ref{fig:setup}a(i), (ii).  The truncated conical shell is installed in an acrylic tube with an inner diameter (25.4 mm) matched to that of the cone base, constitutes a valve testing module as shown in Fig.~\ref{fig:setup}b. Two holes were bored into the tube wall to measure the pressure difference across the cone.  The tube is connected to a chamber with viewing windows where a camera can be attached.

\begin{figure}[htbp]
\includegraphics[width=0.8\textwidth]{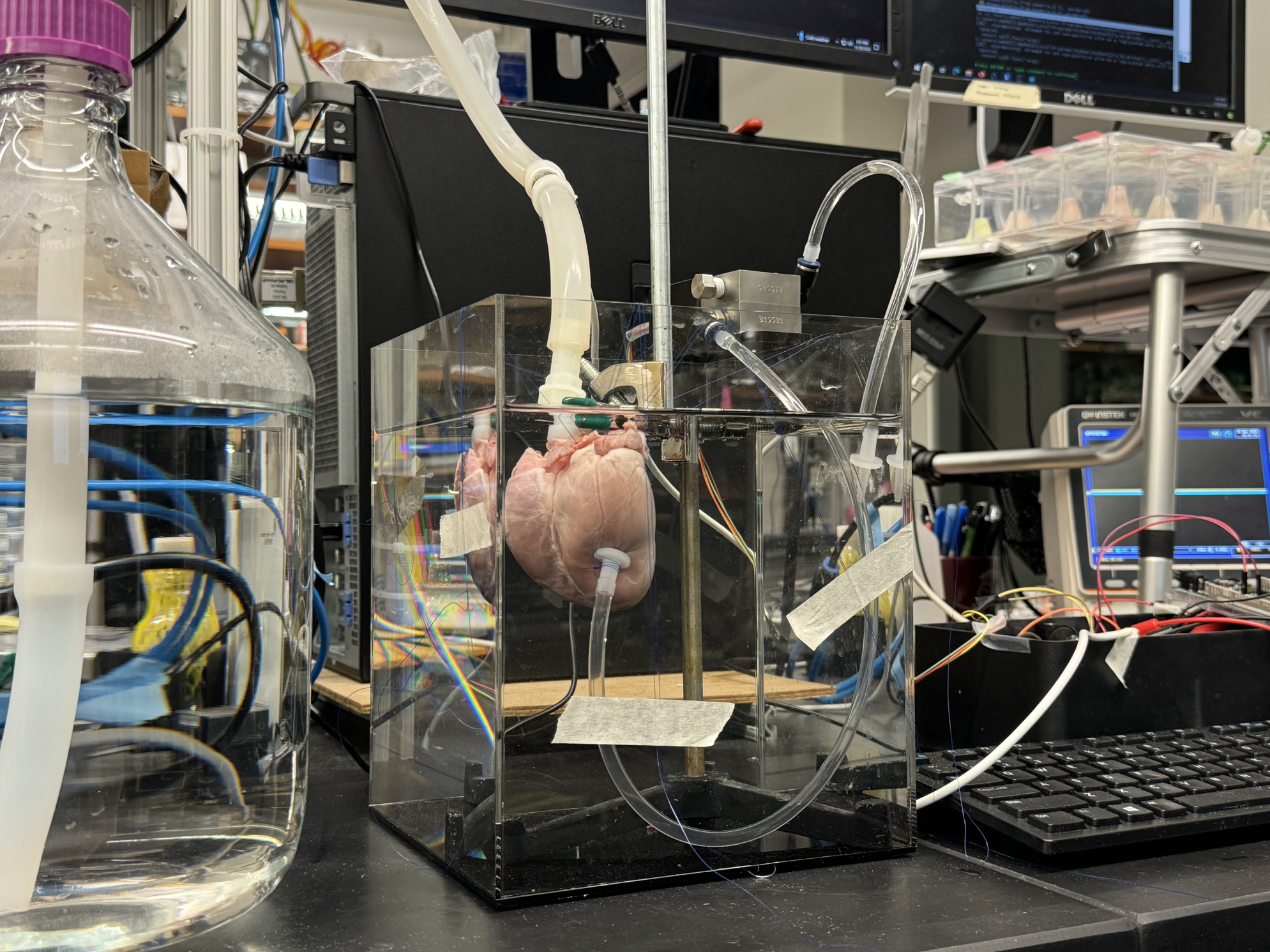}
\caption{Setup to measure the pressure, flow, and geometry of the mitral valve of a Yorkshire pig heart.}
\label{fig:heart}
\end{figure}

Figure~\ref{fig:setup}c shows an overview of the setup to measure the $p$-$Q$ curve of the conical shell.  Regulated by a pressure controller (Elveflow OB1 MK4, \circledno{1}), de-ionized water from two 4L-reservoirs (\circledno{2}) flows through the valve testing module (\circledno{3}) in either direction.
The flow rate $Q$ is measured by a turbine flow transmitter (Vision Turbine Meters BV1000, \circledno{4}).  We use a differential pressure transducer (Validyne P55D, \circledno{5}) to measure the pressure difference $p \equiv p_1-p_2$ across the conical shell. 
The signals are amplified by an non-inverting operational amplifier circuit (AD711JN, \circledno{6}) before being sampled by a microcontroller board (Arduino Due).  
The valve system is illuminated from the side (Sunbeam LED, \circledno{7}).
A DSLR (Nikon D5300,
\circledno{8}) or a high-speed camera (Phantom v9.1)
records the cone morphology from its base side. 
External perturbations are manually injected by tapping a 50 mL syringe (\circledno{9}) connected from the side to the valve testing module.

\begin{figure}[htbp]
\includegraphics[width=0.9\textwidth]{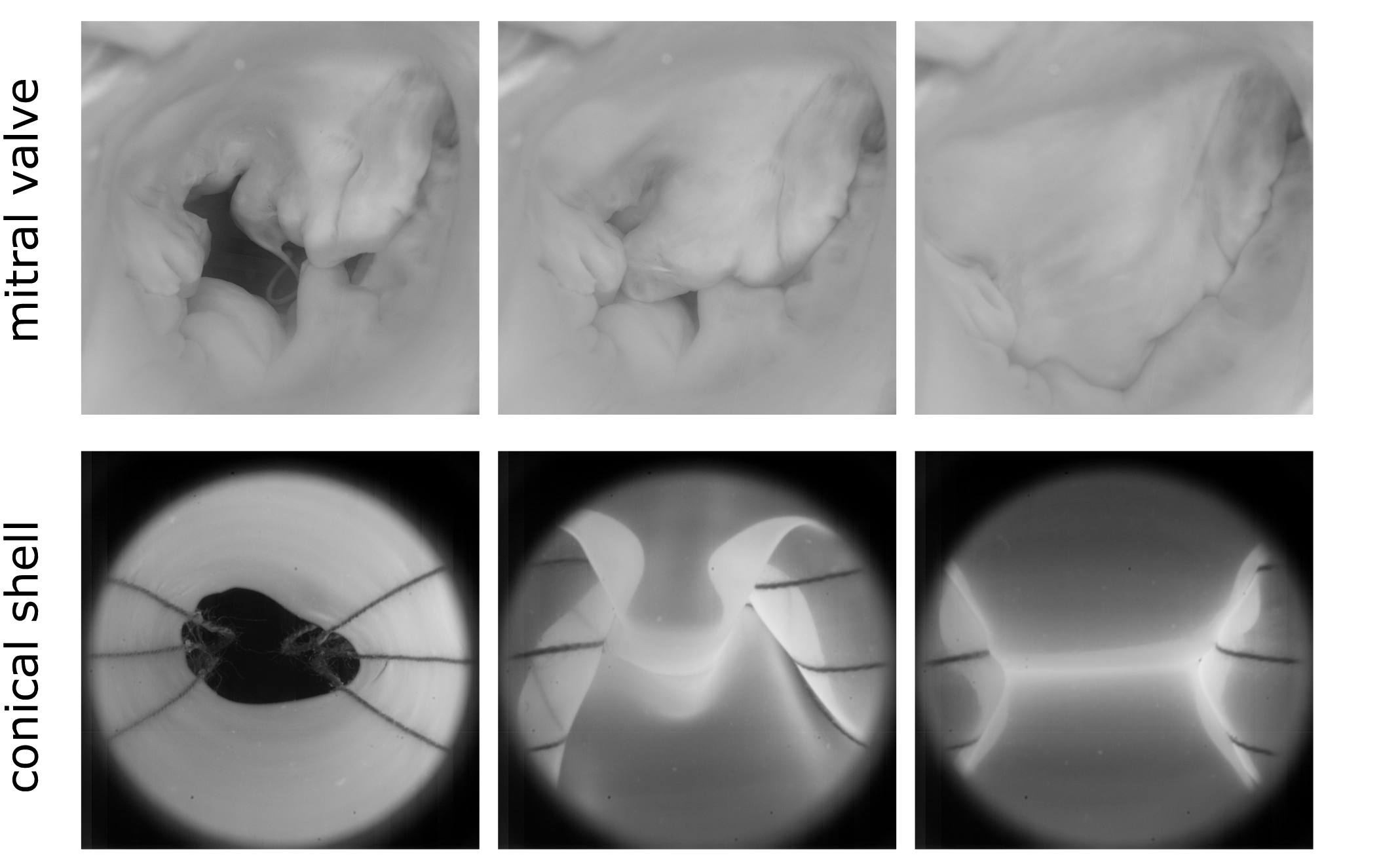}
\caption{Top row: Closure process for the mitral valve of a Yorkshire pig heart under an impinging flow rate of $8.5$ mL/s. Bottom row: Closure process for a tethered conical shell ($\alpha=20^{\circ}$) under an impinging flow which is gradually increased from 0 to $3$ mL/s.}
\label{fig:closure}
\end{figure}

To measure the orifice configuration from a biological mitral valve, we excised the left atrium of a heart from a Yorkshire pig, exposing its atrioventricular junction.  The aorta is tightly fastened to a barbed hose fitting, where an impinging flow is supplied.  The pig heart is then immersed in water (Fig.~\ref{fig:heart}).  One probe of the pressure transducer is pierced through the myocardium into the left ventricle to measure the ventricular pressure, while another probe is placed near the atrioventricular junction from the outside. A high-speed camera (not shown) is placed at the top, pointing downward to record the valve front-view configuration. High-speed image sequences in Fig.~\ref{fig:closure} show a parallel comparison between the valve closure processes for both the conical shell and the mitral valve under an impinging flow.

\begin{figure}[htbp]
\includegraphics[width=0.7\textwidth]{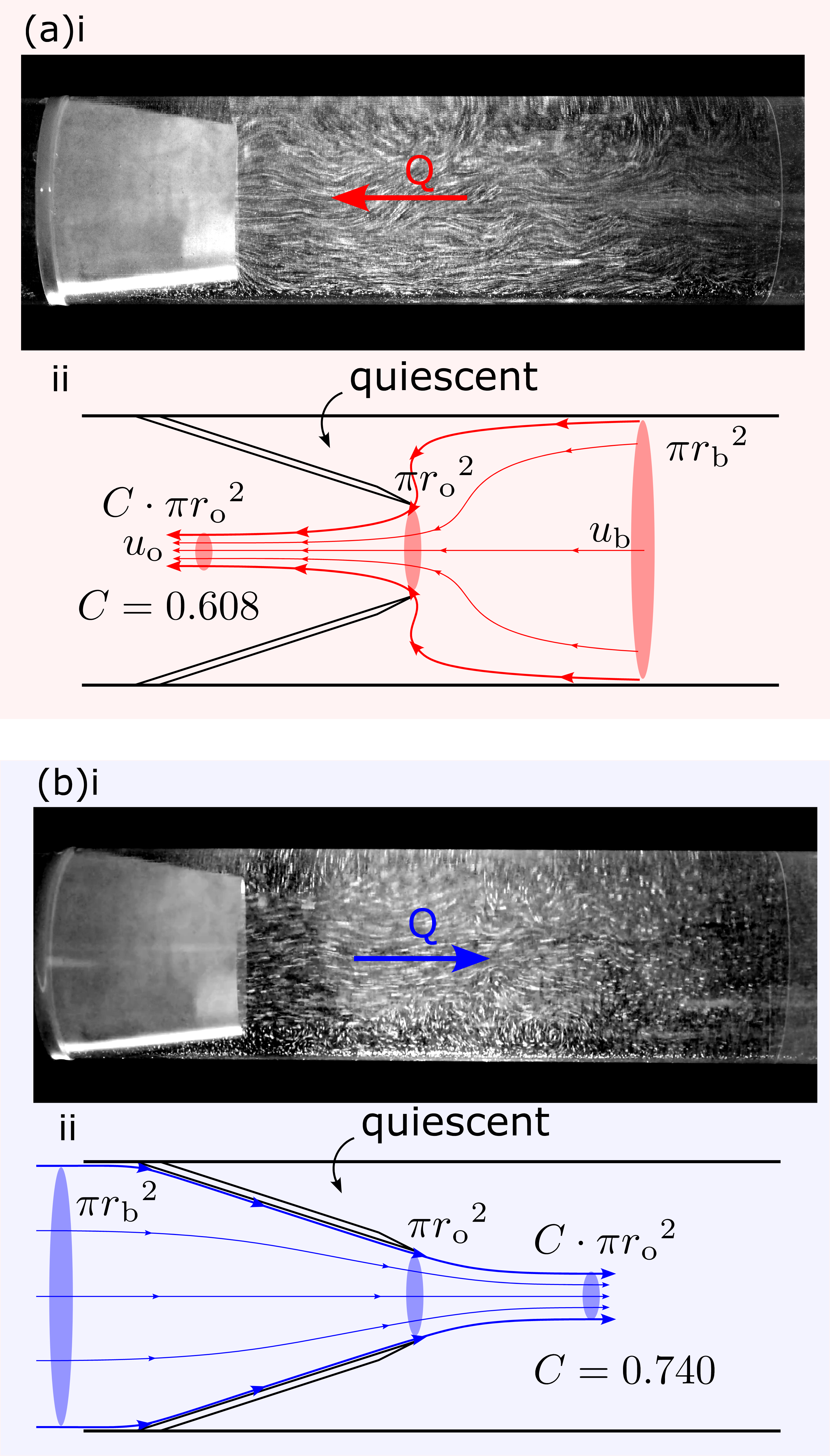}
\caption{(a)i, (b)i The velocity field visualized by seeding particles for a flow against or along the direction of narrowing of a conical shell.  (a)ii, (b)ii Schematics of the corresponding flow field of an ideal fluid.  The regions outside the areas bounded by the thick curves (blue or red) are assumed to be quiescent.}
\label{fig:flowfield}
\end{figure}

\begin{figure}[htbp]
\includegraphics[width=1\textwidth]{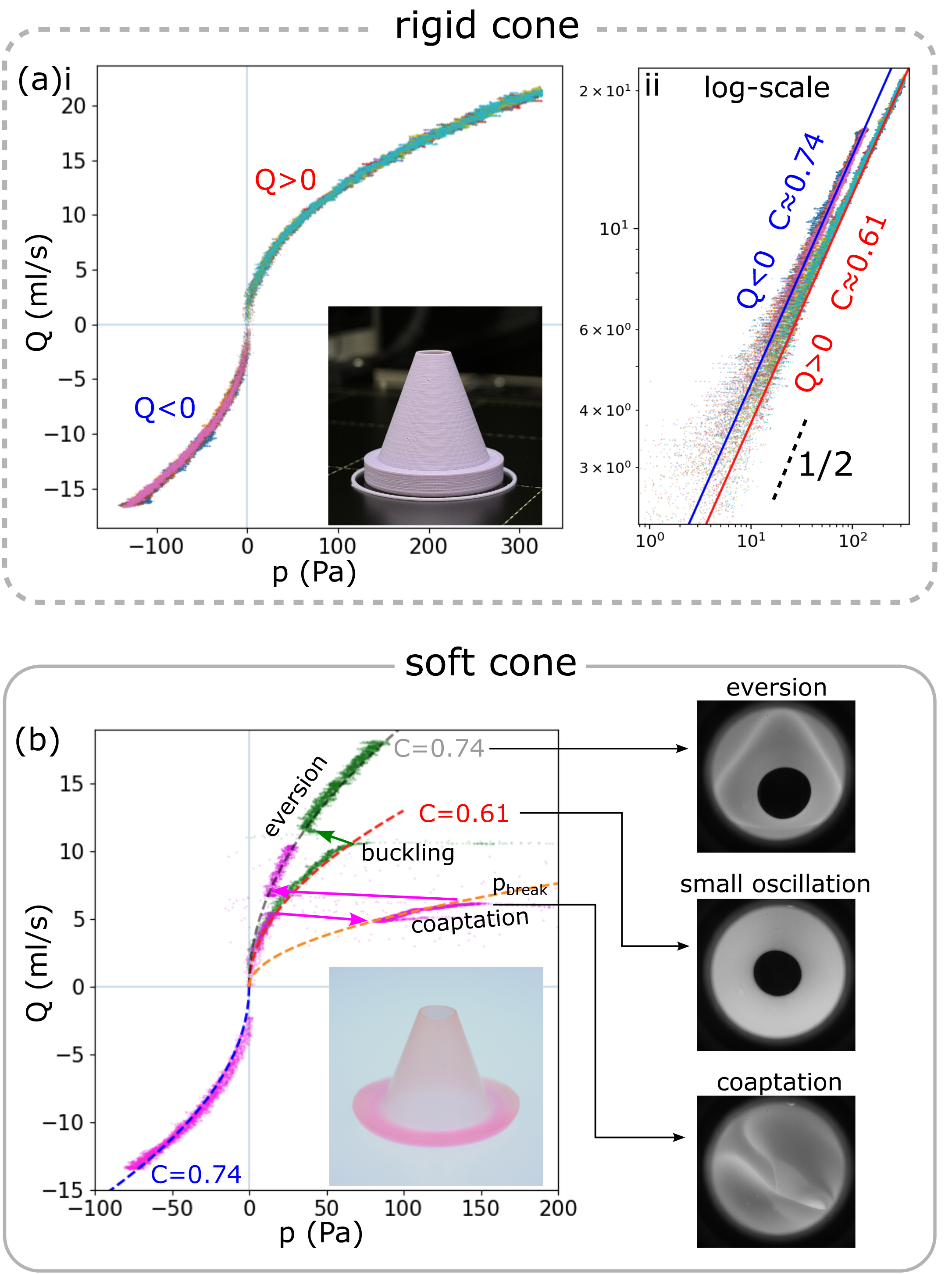}
\caption{(a)i Pressure-flux $p$-$Q$ measured from a rigid cone of half-opening angle $20^{\circ}$.  ii Logarithmic scale of the same data as in (i). The coefficient of contraction for the two flow directions is fitted to be $C=0.74$ (blue line) and $C=0.61$ (red line). (b) $p$-$Q$ curve for a soft, polymeric cone of the same shape as in (a), where $C=0.61$ and $C=0.74$ measured from the rigid cone give good agreement with data from a soft cone (red and blue curves). For the state of coaptation, using $C\cdot S$ as a fitting parameters gives a good match to the data trend (orange dashed curve).}
\label{fig:C}
\end{figure}

\section{Pressure-flow ($p$-$Q$) relation for a deformable cone}
For the flow in the testing module with a conical valve, we visualize the velocity field using seeding particles (50 $\mu$m Polyamide seeding particles, Dantec Dynamics).  Figure~\ref{fig:flowfield}a(i), b(i) show two snapshots of the particle streaks of the opposite flow directions.  These different configurations demonstrate the asymmetry of the flow upon switching the flow direction: whether the flow is along or against the direction of narrowing of the cone,  the fluid movement is always \emph{contractive}.  We sketch the corresponding flow configurations in Fig.~\ref{fig:flowfield}a(ii), b(ii), respectively. We neglect the weak turbulent fluctuations, and assume that the fluid outside the regions confined by the thicker, arrowed lines are approximately quiescent (``dead water''). The Reynolds number associated with these experiment is $\mathrm{Re}\approx500$, which justifies the idealization of an inviscid flow. In our system, for either along or against the direction of narrowing of the cone, we can consider a flow of density $\rho$ passing through two circular cross sections of radii $r\ts{b}$, $r\ts{o}$ with velocities $u\ts{b}$, $u\ts{o}$ under a pressure drop of $p$.  Bernoulli's principle and the conservation of flux lead to:
\begin{align}
\frac{p}{\rho} &= \frac{u\ts{o}^2-u\ts{b}^2}{2}\\
C u\ts{o} r\ts{o}^2 &= u\ts{b} r\ts{b}^2, 
\label{eq:bernoulli}
\end{align}
where $C$ is the coefficient of contraction that describes the reduction of the cross-section area of a jet emanating from an orifice~\cite{birkhoffs}. The flow rate across the junction is
\begin{align}
Q &\equiv u\ts{b}\pi r\ts{b}^2\nonumber\\
&= \sqrt{2}\pi\frac{Cr\ts{o}^2}{\sqrt{1-C^2(r\ts{o}/r\ts{b})^4}}\sqrt{\frac{p}{\rho}}\nonumber\\
&\approx \sqrt{2}CS\sqrt{\frac{p}{\rho}},
\label{eq:Q}
\end{align}
Where $S$ is the cross-section area of the cone orifice.  In the last step, we have used the approximation $r\ts{o}^4 \ll r\ts{b}^4$, a simplification that holds for all cones used in our experiments thanks to the 4th power.

Equation~\ref{eq:Q} was formally derived for a \emph{laminar} flow through a conical shell.
To take into account flow fluctuations, we treat $C$ as an effective coefficient: we maintain the intuitive laminar description giving rise to Eq.~\ref{eq:Q}, but absorb all turbulence effects into the coefficient $C$.  
To obtain the numerical value of $C$, we measure the $p$-$Q$ relation in a \emph{rigid} cone (3D printed polylactic acid).
Figure~\ref{fig:C}a(i) shows the $p$-$Q$ curve of a rigid cone of the same geometry as the soft cone used in Fig.~3(b)-(e) of the main text.  The logarithmic scale in Fig.~\ref{fig:C}a(ii) shows a convincing square-root trend predicted by Eq.~\ref{eq:Q}.  We fit from the two data clusters that $C=0.61$ for flow against the direction of cone narrowing ($Q>0$), while $C=0.74$ for flow along the direction of narrowing ($Q<0$).  These values from a rigid cone give good agreements with the corresponding measurements from a \emph{soft} cone, shown in Fig.~\ref{fig:C}b (red solid curve and blue dashed curves).

As the flux $Q>0$ keeps increasing, the soft cone eventually buckles and enters the state of eversion.  In the $p$-$Q$ space, this is manifested as a discontinuous jump, as shown by the green arrow in Fig.~\ref{fig:C}c. Even though the cone has been turned inside out, the $p$-$Q$ curve still follows a square-root relation.  Moreover, in the state of eversion the flow is along the direction of pointing of the cone again.  Therefore, we use $C=0.74$ in Eq.~\ref{eq:Q} under this condition.  Substituting in the measured orifice size $S$ for the eversion state, which is slightly larger than that of the normal state, we reach an excellent agreement with our measured data (gray dashed curve, I quadrant, Fig.~\ref{fig:C}c.  The sharp jump in the $p$-$Q$ trend is therefore fully explained by two factors: i) the coefficient of contraction, $C$, increases from 0.61 to 0.74 as the cone is turned inside out, and ii) the orifice size, $S$, of an everted cone becomes slightly enlarged compared to that of the normal state (compare the first two images to the right of Fig.~\ref{fig:C}c).

In the presence of strong external noise, the cone collapses early in an impinging flow. Instead of flipping to the state of eversion, the cone is crumpled by the flow to a state of coaptation, where small leaking channels form, allowing a small flow (magenta markers, Fig.~\ref{fig:C}c). Although the orifice of the coaptation state cannot be seen, the $p$-$Q$ relation can still be fitted by a square-root curve when we use the combination of $CS$ as a fitting parameter (orange dashed curve, Fig.~\ref{fig:C}c).  When the pressure difference exceeds a break through threshold $p\ts{break}$, the crumpled cone gives way to the flow and enters the state of eversion, analogous to the electric breakdown of a diode circuit.

\section{Stochastic nonlinear oscillator model}
In the conical shell of thickness $e$, length $L$, deformation $h$ and Young's modulus $E$, the azimuthal bending of mode $n$ creates a linear stress $Ee^3r_{ssss}\sim Ee^3n^4h/(2\pi r_o)^4$, where $r(s)$ is the orifice radius as a function of its azimuthal coordinate.  The balance between elasticity and fluid inertia $\rho L h\omega^2$ gives a characteristic frequency~\cite{virot2020s}
\begin{align}
\omega = \frac{1}{(2\pi)^2}\bigg(\frac{E}{\rho}\bigg)^{1/2}\frac{e^{3/2}n^2}{r\ts{o}^2L^{1/2}}.
\label{eq:omega}
\end{align}
To compare with the experiment shown in Fig.~4a of the main text, we substitute in the corresponding cone parameters: $E=0.267$ MPa, $\rho=1$ g/mL, $n=3$, $e=300$ $\mu$m and $L=27$ mm, yielding $\omega\approx2.66$/s.
 
\begin{figure}[htbp]
\includegraphics[width=0.6\textwidth]{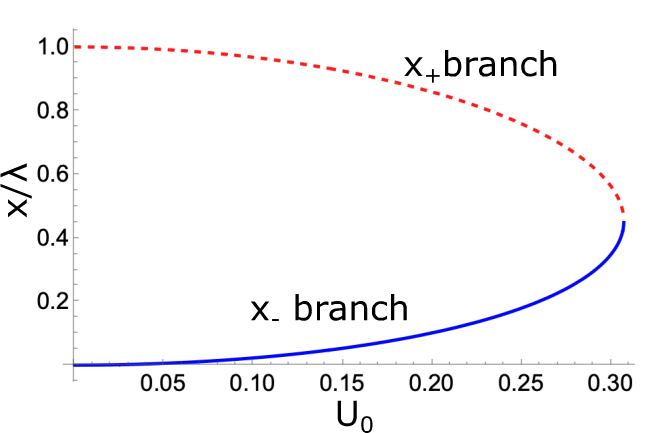}
\caption{Two equilibria, $x_{\pm}$, given by Eq.~\ref{eq:oscillators} with no noise, as a function of $U_0$.  }
\label{fig:saddlenode}
\end{figure}

\begin{figure}[htbp]
\includegraphics[width=0.7\textwidth]{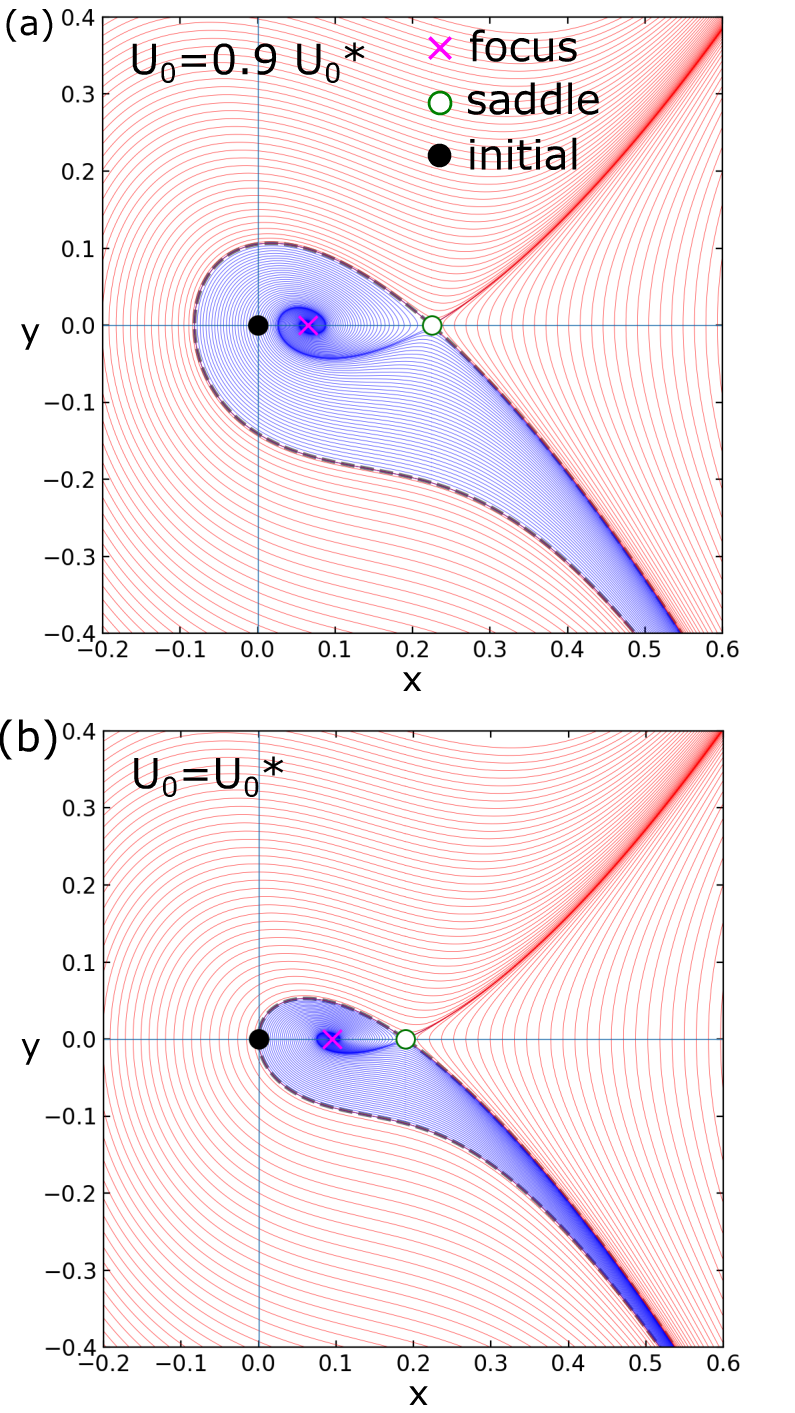}
\caption{Phase portrait of Eq.~\ref{eq:oscillators} with no noise.  (a) When $U_0<U_0^*$, the system's initial condition (black dot) is located within the attracting basin (blue) of the stable node $x_-$ (magenta cross).  (b) When $U_0=U_0^*$, the initial condition sits at the separatrix (dashed curve) that passes the saddle point (green open circle).}
\label{fig:phase}
\end{figure}
Our 1-dimensional elastohydrodynamic model (Eq.~3 of the main text) constitutes a 2nd order nonlinear dynamical system, with the normalized form:
\begin{align}
\frac{\mathrm{d}x}{\mathrm{d}\tau}&\equiv y\nonumber\\
\frac{\mathrm{d}y}{\mathrm{d}\tau}&=(U-2\zeta)+\frac {1}{\lambda}x^2 - (1-U^2)x + U^2\overline{\alpha},
\label{eq:oscillators}
\end{align}
where the we decompose the flow into a mean flow with noise $U = U_0 + \Xi$.    
In the absence of noise $\Xi=0$, Eq.~\ref{eq:oscillators} gives two equilibria 
\begin{align}
x_{\pm}=(\lambda/2)[1-U_0^2\pm\sqrt{(1-U_0^2)^2-4U_0^2\overline{\alpha}/\lambda}],
\end{align} 
where $x_-$ is a stable node, and $x_+$ is a saddle point.  Figure~\ref{fig:saddlenode} (with parameters specified below) shows that the two equilibria, $x_{\pm}$, are created through a saddle-node bifurcation when $U_0$ is lowered below a critical value 0.31.  We neglect two negative branches at large $U_0$.

\begin{figure*}[htbp]
\includegraphics[width=0.95\textwidth]{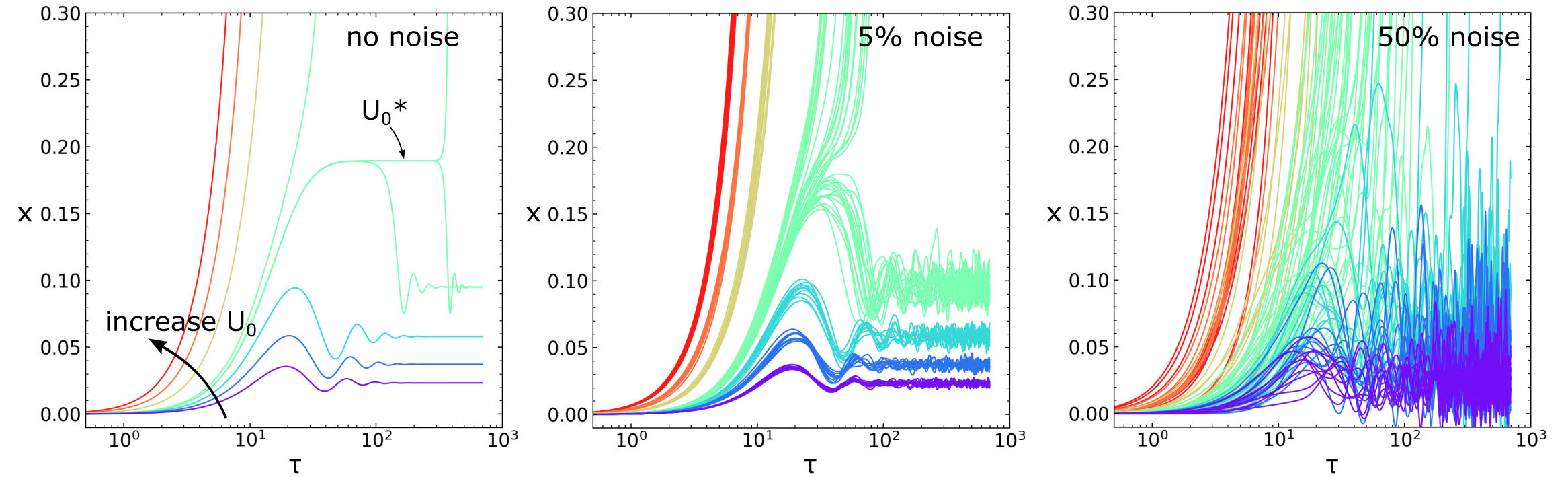}
\caption{Numerical integration of Eq.~\ref{eq:oscillators} with $\sqrt{\langle\Xi^2\rangle}/U_0=0$, 0.05, and 0.5, with parameters $(\overline{\alpha}, \zeta, \lambda) = (0.67, 0.28, 0.31)$ for a range of $U_0$. Each trajectory is repeated 10 times to show the noise effect.  With no noise, the system approaches the saddle point and is restored to the stable node later and later as $U_0\rightarrow U_0^*$.  A small noise broadens each $x$-$\tau$ trajectory into a band.  A large persistent noise triggers early bifurcation so that the originally stable trajectories also diverge.}
\label{fig:xt}
\end{figure*}

The oscillator Eq.~\ref{eq:oscillators} has five control parameters: $\overline{\alpha}$,  $U_0$, $\Xi$, $\zeta$, and $\lambda$.  The first three parameters can be converted from experimental parameters of $\alpha$, $u_0$ and $\xi$.  We obtain the remaining two parameters, $\zeta$ and $\lambda$, in the following way.  For the cone used in our experiment ($\alpha=10^{\circ}$), we first set $u_0 \rightarrow u_0*$ so that the system is near the buckling transition (cyan, Fig.~4 of the main text).  We measure from our experimental curve of $x(t)$ the ratio $x\ts{+, exp}/x\ts{-, exp}=0.18$, where $x\ts{+, exp}$ is the size of the initial hump, and $x\ts{-, exp}$ the final oscillation level.  In our model, this translates to the saddle-node ratio
\begin{align}
 R(U_0^*, \lambda)&\equiv\frac{x_+}{x_-}\nonumber\\
 &=\frac{1-U_0^{*2}+\sqrt{(1-U_0^{*2})^2-4U_0^{*2}\overline{\alpha}/\lambda}}{1-U_0^{*2}-\sqrt{(1-U_0^{*2})^2-4U_0^{*2}\overline{\alpha}/\lambda}}
\end{align}  
We measure the threshold velocity for the buckling transition to be $u_0^*= 2$ cm/s, which translate to $U_0^*=0.29$.  Solving $R(U_0^*=0.29, \lambda) = 0.18$ gives $\lambda=0.311$, or $\ell=2$ mm, consistent with the experimental observation that the cone buckles at an amplitude $\ell \lesssim r\ts{o}$.  We then solve $U_0^*(\lambda=0.311,\zeta)=0.29$ numerically to get $\zeta=0.28$, achieved by adjusting $\zeta$ to enforce the system initial condition onto the \emph{separatrix} when $U_0 = 0.29$.  This process is illustrated in the phase portrait of  Fig.~\ref{fig:phase}.

\begin{figure}[htbp]
\includegraphics[width=1\textwidth]{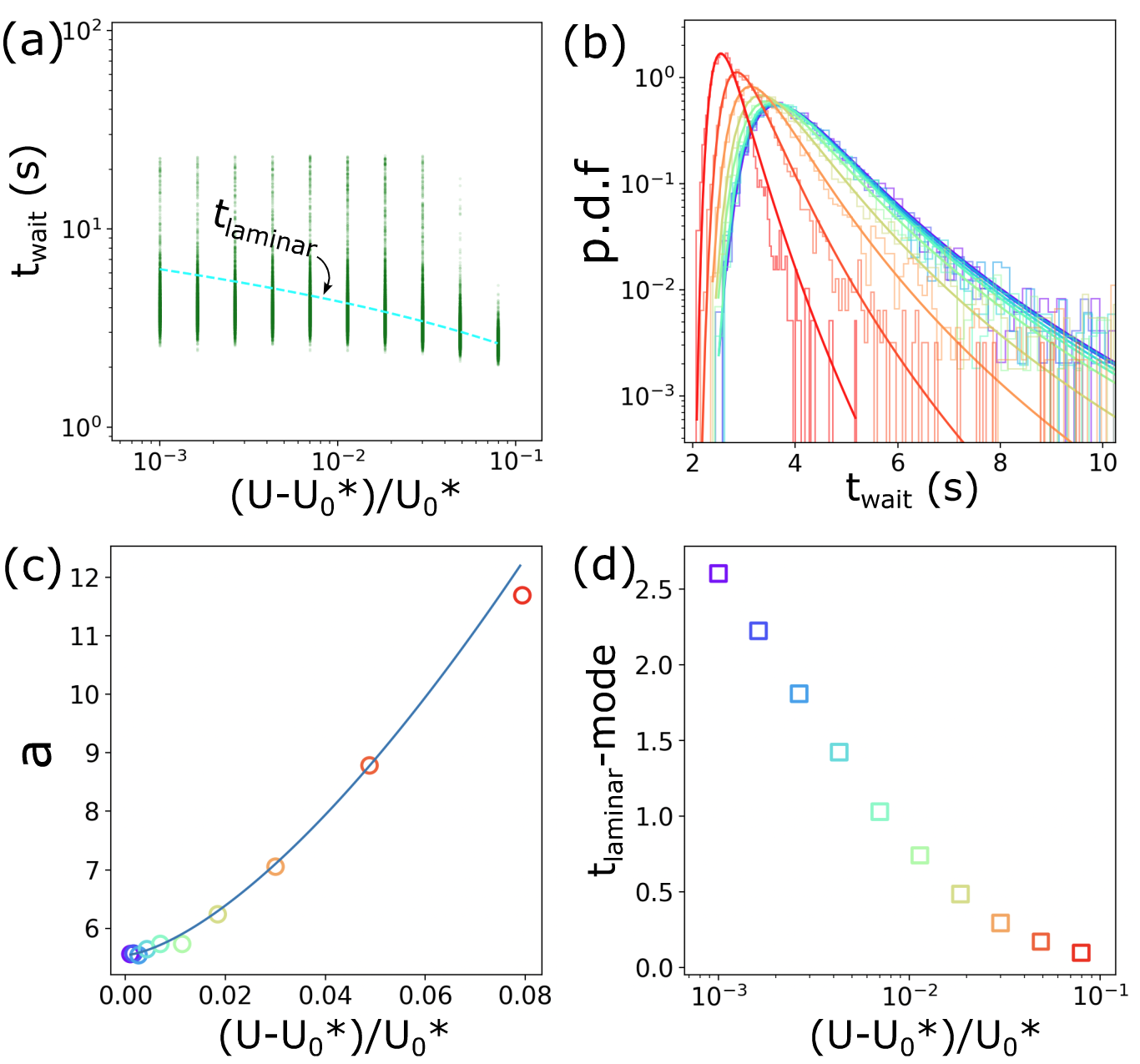}
\caption{(a) The waiting time, $t\ts{wait}$, for the oscillator to diverge, as a function of the distance to the threshold velocity $(U-U_0^*)/U_0^*$. The model integration was repeated by 5000 times with $\sqrt{\langle\Xi^2\rangle}/U_0=0.1$. (b) The histogram of $t\ts{wait}$ given by (a) for each $U_0$ (steps).  The Fr\'echet distribution (curves, Eq.~\ref{eq:frechet}) gives excellent agreement with the simulated results. (c) Fitted parameter $a$ of the Fr\'echet distribution as a function of the distance to the threshold, $(U_0-U_0^*)/U_0^*$. Blue curve, $(a-a^*)\propto[(U_0-U_0^*)/U_0^*]^{3/2}$ where $a^*\approx5.6$. (d) Deviation of the mode, $s(a/(1+a))^{1/a}$, from the waiting time given by the model with no noise, $t\ts{laminar}$, as a function of $(U_0-U_0^*)/U_0^*$.}
\label{fig:frechet}
\end{figure}
With all the parameters, Eq.~\ref{eq:oscillators} is then integrated using Runge-Kutta, for a duration of $20\times2\pi/\omega$ with a time step $\Delta t \ll e/u_0$, leading to the time evolution of $x(\tau)$ shown in Fig.~\ref{fig:xt}a. Figure~\ref{fig:xt}b, c show that adding a small noise $\sqrt{\langle\Xi^2\rangle}/U_0=0.05$ broadens each trajectory into a band, while a large noise $\sqrt{\langle\Xi^2\rangle}/U_0=0.5$ makes originally stable trajectories diverge.

To examine the distribution of the waiting time $t\ts{wait}$, for the oscillator to diverge, we run the model simulation 5000 times for each flow rates, $U_0$, at a 10\% noise, collected in Fig.~\ref{fig:frechet}a.   Figure~\ref{fig:frechet}b shows the histogram for each $U_0$.  The Fr\'echet distribution ~\cite{gumbels}(curves in Fig.~\ref{fig:frechet}b) gives an excellent fit for the long-tailed histograms at all $U_0$:
\begin{align}
f(t; a, s) = \frac{a}{s}\bigg(\frac{t}{s}\bigg)^{-1-a}e^{(\frac{t}{s})^{-a}},
\label{eq:frechet}
\end{align}
where $a$ is the shape parameter, $s$ is the scale factor, and the mode of the distribution is given by $s(a/(1+a))^{1/a}$ ($\approx s$ in our case). The fitted values of $a$, $s$ are shown in Fig.~\ref{fig:frechet}c, d, demonstrating that as $U_0\rightarrow U_0^*$, the width of the distribution plateaus to $a^*\approx5.6$ with a 1.5-power law (blue curve, Fig.~\ref{fig:frechet}c) while the mode shifts logarithmically (Fig.~\ref{fig:frechet}c) with respect to $t\ts{laminar}$, the waiting time given by a flow with no noise.

\section{Triggering the buckling transition with a deterministic disturbance}

\begin{figure}[htbp]
\includegraphics[width=0.8\textwidth]{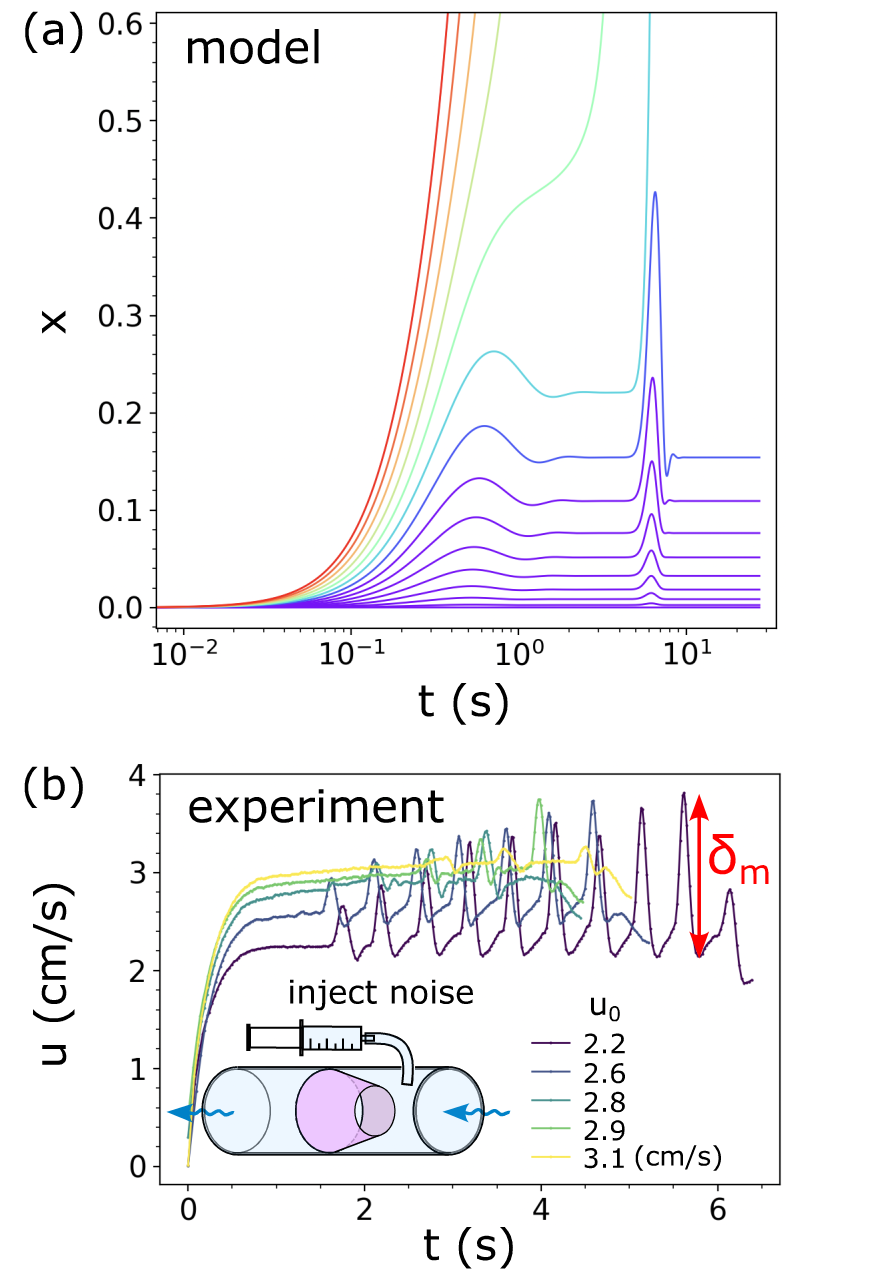}
\caption{(a) Time evolution of the solution of the nonlinear oscillator model Eq.~\ref{eq:oscillators} ($\alpha=20^{\circ}$) with a single spike of disturbance of amplitude $\Delta=0.3U_0$ at $t=6$ s. (b) Measured flow velocities $u = u_0 + \delta$ for various base flows $u_0$, each with a sequence of injected noise spikes of an increasing amplitude until the cone collapses at $\delta\ts{m}$. Schematic: the experimental setup to inject controlled noise spikes.}
\label{fig:spikess}
\end{figure}

Our 1-dimensional stochastic nonlinear oscillator model offers an explanation for the early buckling of the cone under a flow, demonstrated in Fig.~\ref{fig:C}c above (also see Fig.~3b in the main text).  
 We note that the node $x_{-}$ is stable to a steady flow, a large disturbance can push the system beyond the saddle  $x_{+}$, even when $U_0<U_0^*$; indeed the finite perturbation amplitude required to drive the system out of the attracting basin of $x_{-}$ is a characteristic signature of a subcritical bifurcation.  Figure~\ref{fig:spikess}a shows the effect of a single noise spike to a $\alpha=20^{\circ}$ cone for various (dimensionless) base flow velocities. The spike amplitude is set to be $\Delta=0.3U_0$ and is inflicted at $t=6$.  As a result the system diverges at a significantly lower $U_0^*\approx0.28$ compared to the threshold $U_0^*\approx 0.32$ given by the same oscillator with no noise.

Correspondingly, in our experiment, we use a syringe connected to the front of a conical shell of $\alpha=20^{\circ}$ (schematic of Fig.~\ref{fig:spikess}b) to inject a sequence of spikes of progressively larger amplitude, until the cone collapses.  The time interval between spikes is set to be sufficiently long to avoid cumulative effects.  Figure~\ref{fig:spikess}b shows the measured (dimensional) flow velocities including the spike, $u=u_0+\delta$, until the cone collapses, for various base flows $u_0$.  The last spike recorded in each curve reflects the minimum disturbance $\delta\ts{m}$ that trigger the buckling transition.  These measured spikes are plotted in the (normalized) $U_0$-$\Delta$ space in Fig.~5 of the main text, demonstrating that our model gives a quantitative explanation for the early transition of our conical shell under a disturbance, thus facilitating flow rectification.

\end{document}